\documentclass[aps,pre,reprint,twocolumn,superscriptaddress,showpacs,floatfix]{revtex4-1}
\usepackage[latin1]{inputenc}
\usepackage{epsfig}
\usepackage{amsfonts}
\usepackage{amssymb}
\usepackage{calrsfs}
\usepackage{amsmath}
\usepackage{xcolor}
\usepackage{graphicx}
\usepackage[colorlinks=true,allcolors=blue]{hyperref}
\hypersetup{breaklinks=true}

\begin{document}

\title{Periodically driven harmonic Langevin systems}
\author{Shakul Awasthi}
\email{shakulawasthi010615@iisertvm.ac.in}
\author{Sreedhar B. Dutta}
\email{sbdutta@iisertvm.ac.in}
\affiliation{School of Physics, Indian Institute of Science Education and Research Thiruvananthapuram, Thiruvananthapuram 695551, India}

\date{\today}

\begin{abstract} 
Motivated to understand the asymptotic behavior of periodically driven thermodynamic systems, we study the prototypical example of Brownian particle, overdamped and underdamped, in harmonic potentials subjected to periodic driving. The harmonic strength and the coefficients of drift and diffusion are all taken to be $T$-periodic. We obtain the asymptotic distributions almost exactly treating driving nonperturbatively. In the underdamped case, we exploit the underlying $SL_2$~symmetry to obtain the asymptotic state, and study the dynamics and fluctuations of energies and entropy. We further obtain the two-time correlation functions, and investigate the responses to drift and diffusion perturbations in the presence of driving.  
\end{abstract}

\maketitle

\newcommand{\bee}{\begin{equation}}
\newcommand{\eee}{\end{equation}}
\newcommand{\tm}{(t)}
\newcommand{\gm}{\gamma}
\newcommand{\dbar}{d\hspace*{-0.08em}\bar{}\hspace*{0.1em}}
\newcommand{\haf}{\frac{1}{2}}
\newcommand{\x}{\mathbf{x}}

%
%
%
%

\section{Introduction}

Systems with a large number of degrees of freedom can exhibit a variety of nonequilibrium behavior depending on the imposed macroscopic constraints. Subjecting such systems to periodic driving in time may lead them to states with properties that are not only far from equilibrium but also far from well-explored nonequilibrium states such as, for instance, local-equilibrium states or steady states. Periodically driven systems under suitable conditions can exhibit oscillating behavior that may be required to be understood in its own right and not as some perturbative piecemeal extension of our understanding of other nonequilibrium states.

  Extensive studies long have been done on periodically driven systems. There has been an escalating interest in recent times on the effect of periodic driving in a variety of systems, including classical dynamical systems~\cite{Salerno2016, Higashikawa2018}, quantum systems~\cite{Bukov2015, Eckardt2015}, and stochastic systems~\cite{Jung1993, Kim2010, Kohler1997, Wang2015, Knoch2019, Brandner2015}. Various studies were done many decades ago in the context of stochastic resonance~\cite{Gammaitoni1998}, where it was customary to consider periodic driving with small amplitude. There were also many studies done, if not by imposing any restriction on the amplitude, then by assuming the timescale of driving to be either small or large compared to the relevant timescales associated with the time-independent forces~\cite{Bukov2015, Dutta2003}  These studies, or perturbative extensions of them, may not be adequate to describe the state or the thermodynamic properties of periodically driven many-particle systems.
 
 Periodic driving presumably can lead a macroscopic system to some kind of an {\it oscillating state}. If so, then it is natural to ask the following questions in order to effect a description and probe the properties of such states. What are the conditions under which oscillating states can exist? What is the nature of these states, and how different are they from equilibrium and from other nonequilibrium states? What is the minimum set of periodic time-dependent macroscopic variables that is required to parametrize the oscillating states? What is the thermodynamic interpretation of these variables? What are the relevant statistical observables of the oscillating states?

 In order to address these questions, we first need to find the relevant framework to describe the thermodynamic properties of periodically driven systems. Presumably the relevant degrees of freedom are a few macroscopic variables, such as collective variables, which are stochastic and evolve by a continuous Markov process. In this case, the appropriate dynamics is governed by the Langevin, or equivalently, the corresponding Fokker-Planck equation with periodic time-dependent drift and diffusion coefficients. In other words,  the late-time asymptotics of mesoscopic stochastic thermodynamics, wherein periodic driving is appropriately incorporated, could possibly capture the thermodynamic nature of the oscillating state. 

Given this motivation, it may be appropriate to study the prototypical examples of Brownian particle, overdamped and underdamped, subjected to periodic forces. We aim to investigate the effect of periodic driving  on the behavior of various observables. The generic properties presumably can be uncovered by studying exactly solvable models. Hence, by and large, we restrict our study to the harmonic systems as we find them almost exactly solvable even in the presence of driving.  

The outline of the current work is as follows. In the next section, we find almost exactly the asymptotic probability distributions of periodically driven Brownian particle in time-dependent harmonic potentials. In the underdamped case, we exploit the underlying symmetry to obtain the late-time behavior. In Sec.\ref{3TQ}, we investigate the dynamics and fluctuations of various thermodynamic properties, in particular, energies and entropy, in the framework of stochastic thermodynamics. In Sec.\ref{2Tquant}, we determine the two-time correlation functions, and study the response of the stochastic system to drift and diffusion perturbations. Finally, we summarise and briefly conclude in Sec.\ref{conc}.


%
%
%
%

\section{Asymptotic Probability Distribution}\label{2APD}


In this section, we will first introduce a class of stochastic dynamics  that is broad enough to accommodate periodically driven thermodynamic systems. Then we will mainly discuss the prototypical examples of overdamped  and underdamped Brownian motion under the influence of periodically modulated time-dependent forces. 

Presumably it is reasonable to assume that the relevant macroscopic degrees of freedom of a thermodynamic system fluctuate continuously and follow a Markov process.  A continuous Markov process $\mathbf{X}_{t}$ is governed by the set of stochastic equations
\bee \label{lan1}
dX_{t}^{\mu} = f^{\mu}(\mathbf{X}_{t},t)dt + \sigma_a^{\mu}(\mathbf{X}_{t}, t) \cdot dB^a_{t}~,
\eee
where~$\mu$ and~$a$ index the components of~$\mathbf{X}_{t}$ and the standard Weiner process~$dB_{t}$, respectively, while~$f$ and~$\sigma$ are given functions that define the specific process~\cite{Gardiner1985}. The probability distribution~$P(\x,t)$ of the process satisfies the corresponding Fokker-Planck (FP) equation
\bee \label{fp1}
\frac{\partial}{\partial t}P(\x,t) =  \mathcal{L}(\x,t) P(\x,t) ~,
\eee
where the FP operator~$\mathcal{L}$ is of the form
\bee
\mathcal{L}(\x,t) =  -\frac{\partial }{\partial x^{\mu}} F^{\mu}(\x,t) + \frac{1}{2}\frac{\partial^{2}}{\partial x^{\mu} \partial x^{\nu}}D^{\mu\nu}(\x,t)~.
\eee
The drift~$F^{\mu}$ and the diffusion coefficients~$D^{\mu\nu}$ are fixed in terms of the functions~$f$ and~$\sigma$. For Ito interpretation, the drift~${ F^{\mu}(\x,t) = f^{\mu}(\x,t) }$ and the diffusion coefficients~${ D^{\mu\nu}(\x,t)=\sigma_{a}^{\mu}(\x,t)\sigma_{b}^{\nu}(\x,t) \delta^{ab} }$. Here the summation is implied by the repeated indices.

We further assume that the effect of periodic driving is completely captured in the choice of the relevant variables and  in the explicit periodic time dependence of the drift and the diffusion coefficients. In other words, we consider the FP operator to be $T$-periodic, namely ${ \mathcal{L}(\x,t)  = \mathcal{L}(\x,t+T) }$. 

The formal solution of Eq.\eqref{fp1} is given by
\bee\label{form-sol}
P(\x,t) = \mathcal{U}(\x;t,t_0) P(\x,t_0) := \mathcal{T}\big\lbrace e^{\int_{t_0}^{t} \mathcal{L}(\x,t) dt} \big\rbrace  P(\x,t_0) ~,
\eee
where $\mathcal{T}$ denotes the time ordering and~$P(\x,t_0)$ is a given distribution at some initial time~$t_0$. To define the asymptotic distribution unambiguously we will rewrite~${ t= \tau + NT + t_0 }$, where~${ 0 \le \tau < T }$ and~$N$ is an integer, and take the limit~${ N \to \infty }$. We will choose~${ t_0=0 }$, though not required. Note that the evolution operator~$\mathcal{U}$ is invariant under time translation by a period, namely
\bee\label{U-Tperiod}
\mathcal{U}(\x;t + T, t'+T) = \mathcal{U}(\x;t,t')~,
\eee
for any~$t, t'$. This property enables us to write the asymptotic distribution~$P_{\infty}(\x,\tau)$ as
\begin{align}
P_{\infty}(\x,\tau) &:= \lim_{N\to\infty} P (\x, \tau + NT) \nonumber \\
& = \lim_{N\to\infty} \mathcal{U}(\x; \tau, 0) \mathcal{U}(\x; T,0)^N P(\x,0) \nonumber \\
& = \mathcal{U}(\x; \tau, 0) P_{\infty}(\x,0)~,
\end{align}
where~$P_{\infty}(\x,0)$ satisfies the condition
\bee
\mathcal{U}(\x; T,0)P_{\infty}(\x,0)=P_{\infty}(\x,0)~.
\eee
A necessary condition for the existence of~$P_{\infty}(\x,0)$ is that the real part of none of the eigenvalues of the operator~$\mathcal{U}(\x;T,0)$ exceeds unity. Further, it does not depend on the initial condition provided the operator has a unique normalizable eigenfunction with eigenvalue one that is separated from the nearest eigenvalue with a gap. Note that the asymptotic distribution satisfies the periodicity condition:~${ P_{\infty}(\x,\tau+T)=P_{\infty}(\x,\tau) }$. We will refer to the state associated with this periodic asymptotic distribution~$P_{\infty}(\x,t)$ as an oscillating state and relabel the distribution as~$P_{os}(\x,t)$.

When the time-period of driving is comparable to other timescales in the problem, then the perturbative treatment of driving may not be sufficient to deduce the nonequilibrium features of the oscillating state. The extent to which the perturbative treatment deviates from an exact solution can be estimated even within perturbation theory. For instance, consider the familiar example of classical harmonic oscillator, extended by a simple time-dependent perturbation, governed by the equation $\ddot{x} + \omega_0^2 x = \omega_1^2 \cos(\omega t) x$.  When~$\omega_1^2 \ll \omega_0^2$, we could compromise with a perturbative solution~$x=x_0+x_1+\cdots$, where $x_0= a \cos(\omega (t-t_0))$ and the first order correction goes as $x_1 \sim \omega_1^2 \omega^{-1} (2\omega_0 \pm \omega)^{-1} a$. Now the first-order term and the rest of the corrections begin to dominate when~$\omega \sim 2 \omega_0$, and the perturbative solution cannot be trusted anymore. Similarly, in case of the FP equation, suppose we decompose the FP operator~$\mathcal{L} = \overline{\mathcal{L}} + \Delta\mathcal{L}$, where~$\overline{\mathcal{L}}$ is the FP operator averaged over the time-period~$T$ and~$\Delta\mathcal{L}$ denotes the purely oscillating part. We could treat the time-dependent part perturbatively, when the typical scale of~$\overline{\Delta\mathcal{L}^2}$ is much smaller than the typical scale of~$\overline{\mathcal{L}}^2 $. The perturbative corrections will contain factors of the form~$\parallel \overline{\Delta\mathcal{L}^2}\parallel^{1/2} / \left( T^{-1} -\parallel\overline{\mathcal{L}} \parallel\right)$, where~$\parallel L\parallel$ denotes any one of the scales associated to the operator~$L$. Essentially, when~$T \parallel\overline{\mathcal{L}} \parallel \sim 1$, the perturbative analysis cannot capture the properties of the oscillating state.  This clearly provides us the motivation to obtain the asymptotic distributions exactly, at least in some special cases, and study their features. Hence we proceed in the remainder to investigate these prototypical examples restricted to the harmonic cases. We shall refer those cases as harmonic for which the diffusion coefficients~${ D^{\mu\nu}=D^{\mu\nu}(t) }$ are independent of~$\x$ and the drift coefficients take the form~${ F^{\mu}= A^{\mu~}_{~\nu}(t)x^{\nu}+b^{\mu}(t) }$, where the functions~$A$ and~$b$ can depend only on~$t$.

%
%
%
%

\subsection{Overdamped oscillator}

We shall first consider a one-dimensional overdamped Brownian particle in driven harmonic potential following the trajectory~$X_t$ described by the Langevin equation
\bee
\gm \tm \dot{X}_{t}  = - k \tm X_{t}  +\eta \tm ~.
\eee
The noise~$\eta$ is Gaussian with zero mean and periodic time-dependent variance, namely $\langle \eta\tm \rangle_{\eta} = 0$ and $\langle \eta\tm \eta(t') \rangle_{\eta}= 2D\tm \delta(t-t')$. The parameters~$\gm, D$ and~$k$ are taken to be $T$-periodic.
Since the viscosity and the noise together model the interaction with the environment, it can be expected that even without any fine tuning the parameters~$\gm$ and~$D$ carry the same periodicity.  On the other hand, since the external forces can be driven independently, the choice of~$k$ having the same periodicity is more a restriction than a rule. On physical grounds we may require~$\gm$ and~$D$ to be positive at any time~$t$, while such a restriction, as we shall show, can be weakly relaxed for~$k$ during some time. 

The probability distribution~$P(x,t)$ of~$X_t$ satisfies the corresponding FP equation 
\bee
\frac{\partial}{\partial t} P(x,t)=  \left[ \frac{k \tm }{\gm \tm }\frac{\partial}{\partial x} x + \frac{D\tm }{\gm ^{2}\tm }\frac{\partial^2}{\partial x ^{2}} \right]P(x,t)~,
\eee
and hence its moments~$X_n := \int dx~ x^n P(x,t)$, for natural boundary conditions, follow the equations
\bee\label{Mn-eqn}
\frac{d}{dt}X_{n}\tm  = - n \frac{k \tm }{\gm \tm }X_{n}\tm  + n(n-1)\frac{D\tm }{\gm ^2\tm }X_{n-2}\tm~. 
\eee

The fact that the  asymptotic distribution under certain conditions is Gaussian can also be deduced from the asymptotics of the moments.
The first moment~$X_1\tm$ evolves from its initial value~$X_1(0)$ as follows:
\bee
X_{1}\tm  = X_{1}(0)\exp{ \left[-\int_{0}^{t}dt'~\frac{k (t') }{ \gm(t') } \right]}~,
\eee
and vanishes for large time or, equivalently, in the limit~${ N\to\infty }$, provided 
\bee\label{cond-k>0}
\overline{\left( k/ \gm \right)} := \frac{1}{T} \int_0^T dt' \left(\frac{k(t')}{\gm(t')}\right) > 0~.
\eee
The second moment 
\bee\label{Sol-2m}
X_{2}\tm  = X_{2}(0) e^{-2\int_{0}^{t} \!\!dt' \frac{k (t')}{\gm (t')}} + \int_{0}^{t} \! \!dt' \frac{2D(t')}{\gm ^{2}(t')}e^{-2\int_{t'}^{t}\!\!dt'' \frac{k (t'')}{\gm (t'')}}~,
\eee
approaches the function~${ \widetilde{X}_2(\tau) = \lim_{N\to\infty} X_2(NT+\tau) }$ asymptotically when the same condition\eqref{cond-k>0} holds, independent of its initial value~$X_2(0)$. It is easy to verify that~$\widetilde{X}_2(\tau)$ is $T$-periodic when it is rewritten as
\bee\label{Asy-2m}
\widetilde{X}_{2}(\tau) = \frac{1}{e^{2T\overline{(k/\gm )} }-1}\kappa(T,\tau) + \kappa(\tau,\tau) ~,
\eee
where 
\bee
\kappa(\tau_{1},\tau_{2}) = \int_{0}^{\tau_{1}} d\tau' \frac{2D(\tau')}{\gm ^{2}(\tau')}\exp{\left[ -2\int_{\tau'}^{\tau_{2}}d\tau'' \frac{k (\tau'')} {\gm (\tau'')}\right]}~.
\eee

We can now deduce the Gaussian nature of the asymptotic moments~$\widetilde{X}_n(\tau)$.  To this end, we define the quantity~${ Y_n \tm=X_{n}\tm - (n-1)X_{2}\tm X_{n-2}\tm }$ and obtain its dynamics, using equation\eqref{Mn-eqn},  as follows:
\bee\label{dif-eq-mo}
\bigg( \frac{d}{dt} + n \frac{k\tm}{\gm\tm}  \bigg) Y_{n}\tm  =  (n-1)(n-2) \frac{D\tm }{\gm(t)^2}  Y_{n-2}\tm~.
\eee
The above equation implies that, when the condition\eqref{cond-k>0} holds, the quantity~$Y_2$ vanishes asymptotically. Further, under the same condition, the quantity~$Y_n$ also vanishes asymptotically provided~${Y_{n-2} =0}$, for any positive integer~$n$. Thus we deduce, by induction, that the property of Gaussian decomposition, ${ \widetilde{X}_{n}(\tau) = (n-1)\widetilde{X}_{2}(\tau)\widetilde{X}_{n-2}(\tau) }$, holds  for all even moments and that all the odd moments vanish.

Hence the asymptotic behavior of the overdamped Brownian particle is governed by the oscillating distribution   
\bee
P_{os}(x,t) = \frac{1}{\sqrt{2\pi \widetilde{X}_{2}\tm }}\;\; \mathrm{exp} \left[-\frac{x^{2}}{2\widetilde{X}_{2}\tm }\right]~.
\eee
We reiterate that for the existence of the oscillating state it is not necessary that the function~$k(t)$ is positive at all times, but rather it is sufficient that the mildly weaker condition\eqref{cond-k>0} holds. In general  it is not required for the FP operator~$\mathcal{L}(\x,t)$ to be negative semidefinite at all times for the eigenvalues of the operator~$\mathcal{U}(\x;T,0)$ to not exceed unity.  It further indicates that periodic driving can indeed enhance the stability of the asymptotic state of the stochastic systems. 

It is useful to note that if the FP equation has a unique solution for a given initial condition, then one can also obtain the asymptotic distribution by choosing the initial condition such that the solution is periodic. For instance, the asymptotic expression\eqref{Asy-2m} can also be obtained from the solution\eqref{Sol-2m} by choosing~$X_2(0)$ such that~${ X_2(\tau+T)=X_2(\tau) }$ instead of taking~${ N\to\infty }$ limit.

%
%
%
%

\subsection{Underdamped oscillator} 


We now extend the analysis of the asymptotic behavior of the periodically driven Brownian particle by relaxing the overdamped limit. One of the motivations for including the inertial term is because with the increase in the frequency of driving the inertial force increasingly dominates over the viscous force. This may bring about nontrivial coupling between velocity and position degrees of freedom. 

The position~$X_t$ and the velocity~$V_t$ of a Brownian particle in a periodic harmonic potential is described by
the set of stochastic equations 
\begin{align}\label{stoc-dyn}
\dot{X}_t& = V_t ~,\nonumber \\
\dot{V}_t& = -\gamma(t)  V_t - k(t) X_t + \eta(t)~,
\end{align} 
where the noise is Gaussian, as specified earlier, with strength~$2D(t)$, and the parameters~$\gm, D$ and~$k$ are $T$-periodic. The corresponding FP equation of the probability distribution~$P(x,v,t)$ is given by
\bee\label{oFP-eqn}
\frac{\partial}{\partial t}P = \bigg[ -\frac{\partial}{\partial x}v - \frac{\partial}{\partial v}\left[ -(\gm  v + k   x)\right]  + D  \frac{\partial^{2}}{\partial v^{2}} \bigg]P~,
\eee
while the moments $ X_{m,n}\tm  := \int dx dv~ x^{m} v^{n} P(x,v,t)$ of the probability distribution, for natural boundary conditions,  satisfy the equations
\begin{align}\label{om2}
\frac{d}{dt}&X_{m,n}\tm  = m X_{m-1,n+1}\tm  - n \gm \tm  X_{m,n}\tm \nonumber  \\  &-  n k \tm  X_{m+1,n-1}\tm  + D\tm  n(n-1) X_{m,n-2}\tm  ~.
\end{align}
Let us classify the moments into various levels, labeled by a positive integer~$L$, wherein the moment~$X_{m,n}$ is said to belong to the level~$L$ when~${m+n=L}$. There are two interesting features of the dynamical equations\eqref{om2} that will almost lead us to its solutions.
One is that they can be viewed, for a given~$L$,  as a set of linear inhomogeneous coupled equations where the homogeneous part contains only level-$L$ moments, while the inhomogeneous part depends only on level-$({L\!-\!2})$ moments.
The second feature is that the homogeneous part of all levels has the same symmetry structure.

\subsubsection{$SL_2$ symmetry}

Suppose we construct a~$L+1$~component vector~${\bf X}_{L} = \left[ X_{L,0},X_{L-1,1}, \cdots, X_{0,L}\right]^T$. Then the homogeneous part of the dynamical equation for~${\bf X}_{L}$ can be written as
\bee\label{L-hom}
\frac{d}{dt}\mathbf{X}^h_{L} =  \left[ \mathbf{J}^{-}_{L} - \frac{\gm \tm }{2} \left( L  I_L +\mathbf{J}_{L} \right) - k \tm  \mathbf{J}^{+}_{L}\right] \mathbf{X}^h_{L}~,
\eee
where~$I_L$ is an identity matrix, and~$\lbrace \mathbf{J}^{\pm}_{L},\mathbf{J}_{L} \rbrace$ are matrices with the following components
\begin{eqnarray}
\left( \mathbf{J}^{+}_{L}  \right)_{r,s}&=& (r-1) \delta_{r,s\!+\!1} ~,\nonumber \\
\left( \mathbf{J}^{-}_{L}  \right)_{r,s}&=& (L+1-r) \delta_{r,s\!-\!1}~, \nonumber  \\
\left( \mathbf{J}_{L}  \right)_{r,s}&=& (2(r-1)-L) \delta_{r,s}~,
\end{eqnarray}
whose indices~${r,s}$ run over~${1,\cdots, (L\!+\!1)}$. It can be verified that these matrices satisfy the commutation relations
\bee\label{sl2alg}
\left[ \mathbf{J}_{L} , \mathbf{J}^{\pm}_{L} \right] = \pm 2\mathbf{J}^{\pm}_{L} ~,\quad  \left[\mathbf{J}^{+}_{L},  \mathbf{J}^{-}_{L} \right] =  \mathbf{J}_{L} ~.
\eee
Hence it follows that the matrices~$\lbrace \mathbf{J}^{\pm}_{L},\mathbf{J}_{L} \rbrace$ are a realization of an~$({L+1})$-dimensional irreducible representation of the generators~$\lbrace \mathbf{J}^{\pm},\mathbf{J}  \rbrace$ of the~$SL_2(\mathbb{R})$ group, respectively. 

To exploit the symmetry without restricting to any particular representation, it is convenient to define the vector~${ \mathbf{Y}_{L}:=\mathbf{X}^h_{L}\exp{\left(L\Gamma/2\right)} }$,  where~${ \Gamma \tm  = \int_{0}^{t}dt' \gm (t') }$, and rewrite equation\eqref{L-hom} as 
\bee\label{L-hom2}
\frac{d}{dt}\mathbf{Y}_{L} =  \left[ \mathbf{J}^{-}_{L} - \frac{\gm \tm }{2} \mathbf{J}_{L}  - k \tm  \mathbf{J}^{+}_{L}\right] \mathbf{Y}_{L}~,
\eee
which has no explicit~$L$ dependence. 

Since any irreducible representation of $sl_2$~algebra is a symmetric power of its standard representation\cite{Fulton2004}, the solutions of level-$L$ homogeneous equation can be obtained by the symmetrization of $L$~tensor powers of the solutions of the level-$1$ equation 
\begin{align}\label{level1-Y}
\frac{d}{dt}
\begin{bmatrix}  
Y_{1,0}\\ 
Y_{0,1}\\ 
\end{bmatrix} = 
\begin{bmatrix}
  \frac{1}{2}\gm\tm&1\\
-k\tm & -\frac{1}{2}\gm\tm 
\end{bmatrix}
\begin{bmatrix}  
Y_{1,0}\\ 
Y_{0,1}\\ 
\end{bmatrix} ~.
\end{align}  
It should be emphasized that this remarkable property of the irreducible representations of $sl_2$ algebra 
plays the most pivotal part to enable us in expressing the higher moments in terms of the lower ones and thus in turn leads us to the asymptotic distribution.

\subsubsection{Level-1 and level-2 moments}

We now show that the level-1 and, hence, level-2 moments can be written down in terms of the solutions of a Hill equation.
This is evident from Eq.\eqref{level1-Y}, from which it follows that the component~$Y_{1,0}$ satisfies the Hill equation 
\bee\label{Hill}
\frac{d^2}{dt^2} Y_{1,0}+ \nu(t) Y_{1,0}=0~,
\eee
where~${ \nu = k- \dot{\gamma} /2 -\gamma^2/4 }$. Let us denote the two independent Floquet solutions  of the Hill equation to be~$u(t)$ and~$w(t)$, satisfying the pseudo-periodic property,~${ u(t+T)=u(t)\exp{(-\mu T)} }$ and~${ w(t+T)=w(t)\exp{(-\mu' T)} }$, respectively. Note that the sum of the constants,~$\mu$ and~$\mu'$, vanishes since~${ u\dot w-w\dot u }$ is conserved. 

The existence of nontrivial solutions and the value of~$\mu$ depends on the function~$\nu(t)$. 
It may be remarked that there are cases for which the solutions are either known~\cite{Casperson1984,Casperson1985,Wu1985}, known to exist~\cite{Brillouin1948}, or known to be stable or bounded under certain conditions~\cite{Magnus2013}.

The two independent solutions of equation\eqref{level1-Y}  can be chosen to be 
\begin{align}\label{sol-Y1}
 \mathbf{Y}_{1}^{(1)} =
 \begin{bmatrix}  
u\\ 
 \dot{u} -\frac{1}{2} \gamma u\\ 
\end{bmatrix} ~, \quad
 \mathbf{Y}_{1}^{(2)} =
 \begin{bmatrix}  
w\\ 
\dot{w} -\frac{1}{2} \gamma w\\ 
\end{bmatrix}~,
\end{align}
 and the corresponding fundamental matrix to be
\bee \label{funda-matrix1}
\mathbf{\Phi}_{1}\tm  = 
\begin{bmatrix}
u &w\\
\dot{u} -\frac{1}{2} \gamma u& \dot{w} -\frac{1}{2} \gamma w
\end{bmatrix} ~.
\eee
In other words,  the solution of Eq.\eqref{level1-Y}, ${ \mathbf{Y}_{1} (t)= \mathbf{\Phi}_{1}\tm \mathbf{\Phi}_{1}^{-1}(0)\mathbf{Y}_{1} (0) }$, can be specified  by the Hill equation solutions~$u(t)$ and~$w(t)$ and the given initial condition~$\mathbf{Y}_{1} (0)$. 

The vector~$\mathbf{X}_{1}$ associated with the level-$1$ moments satisfies the homogeneous equation\eqref{L-hom} and therefore is given by the expression
\bee
\mathbf{X}_{1}\tm =K_1(t,s) \mathbf{X}_1(s)~,
\eee
where the matrix
\bee\label{funda-matrix1}
K_1(t,s)= \mathbf{\Phi}_{1}(t) \mathbf{\Phi}_{1}^{-1}(s) e^{-\left(\Gamma(t)-\Gamma(s)\right)/2 }~.
 \eee
 It can be verified that the fundamental matrix~$\mathbf{\Phi}_1$ satisfies the pseudoperiodic property,~${ \mathbf{\Phi}_{1}(t+T)=\mathbf{\Phi}_{1}(t) \Lambda_1}$, where~$\Lambda_1$ is a diagonal matrix with elements~$\exp(-\mu T)$ and~$\exp(\mu T)$, respectively. Hence it follows that the asymptotic vector
\bee\label{asym-1}
\widetilde{\mathbf{X}}_{1}(\tau) = \lim_{N\to\infty}\mathbf{X}_{1}(NT+\tau) \to 0~,
\eee
provided the modulus of the real part of~$\mu$ and the average~$\overline{\gamma}$ of the viscous coefficient~$\gamma$ over a period, satisfy the condition 
\bee\label{cond-mu-g}
| Re{(\mu)} | < \frac{1}{2} \overline{\gamma}~.
\eee 

The solution~$\mathbf{Y}_{2} (t)$ of Eq.\eqref{L-hom2} for level~$L=2$ can be constructed, as mentioned earlier, from the symmetrized tensor square of the level-$1$ solutions\eqref{sol-Y1}. Thus we obtain the level-$2$ fundamental matrix
\bee \label{funda-matrix2}
\mathbf{\Phi}_{2}\tm  = 
\begin{bmatrix}
u^2 & uw & w^2\\
u U& \frac{1}{2} (Uw+uW)& wW\\
U^2& UW & W^2
\end{bmatrix} ~,
\eee
where~${ U =\dot{u} - \gamma u/2 }$ and~${ W =\dot{w} -\gamma w/2 }$ are used for notational simplicity.
In other words, we have determined the solution~${ \mathbf{Y}_{2} (t)= \mathbf{\Phi}_{2}\tm \mathbf{\Phi}_{2}^{-1}(0)\mathbf{Y}_{2} (0) }$  or, equivalently, obtained the homogeneous part~${ \mathbf{X}^h_{2} (t) = \exp{(-\Gamma(t))}\mathbf{Y}_{2} (t) }$. 

The fundamental matrix~$\mathbf{\Phi}_2$ also satisfies the pseudoperiodic property,~${ \mathbf{\Phi}_{2}(t+T)=\mathbf{\Phi}_{2}(t) \Lambda_2 }$, where~$\Lambda_2$ is a diagonal matrix with elements~${ \exp(-2\mu T), 1, \exp(2\mu T) }$, respectively. This implies that the asymptotic behavior of the homogeneous part is given by~${\mathbf{X}^h_{2} (t) \sim \exp{[-(\Gamma(t)-2t| Re{(\mu)} |)]} }$, and vanishes, in the limit~${t \to \infty}$, when the condition\eqref{cond-mu-g} holds.

Having determined the fundamental matrix\eqref{funda-matrix2}, we can now write down the level-$2$ moments which satisfy the inhomogeneous equation
\begin{align}\label{x2-eqn}
\frac{d}{dt}
\begin{bmatrix}  
X_{2,0}\\ 
X_{1,1}\\ 
X_{0,2}\\ 
\end{bmatrix} = 
\begin{bmatrix}
0&2&0\\
-k & -\gm&1\\
0&-2k & -2\gm
\end{bmatrix}
\begin{bmatrix}  
X_{2,0}\\ 
X_{1,1}\\ 
X_{0,2}\\ 
\end{bmatrix} +
\begin{bmatrix}
0\\0\\2D 
\end{bmatrix}~.
\end{align}  
The general solution of the above equation is given by
\bee\label{2-mom-soln}
\mathbf{X}_2(t)
 = K_2(t,0) \mathbf{X}_2(0)
+ \int_{0}^{t} ds K_2(t,s)\mathbf{b}(s)~,
\eee
where the vector~${ \mathbf{X}_2\tm = \left[X_{2,0}\tm, X_{1,1}\tm, X_{0,2}\tm \right]^T }$, the vector~${ \mathbf{b}\tm = \left[0,0,2D(t) \right]^T }$, and the matrix
\bee\label{ker-matrix2}
K_2(t,s)= \mathbf{\Phi}_{2}(t) \mathbf{\Phi}_{2}^{-1}(s) e^{-\left(\Gamma(t)-\Gamma(s)\right) }~.
 \eee
 We can further read the asymptotic level-$2$ moments from the large time limiting vector~${\widetilde{\mathbf{X}}_2(\tau) = \lim_{N\to\infty} \mathbf{X}_2(NT+\tau)}$. Since the eigenvalues of~${ K_2(T,0)=  \mathbf{\Phi}_{2}(0) \Lambda \mathbf{\Phi}_{2}^{-1}(0) }$ are same as those of~$\Lambda$, in the limit~${N \to \infty}$, the matrix~$K_2(NT,0) \sim \exp[-NT(\bar{\gamma}-2| Re(\mu)|)]$. Thus the loss of the memory of the initial state and the existence of the limiting vector are both ensured when condition\eqref{cond-mu-g} holds. We reiterate that instead of explicitly taking the~$N \to \infty$ limit to obtain~$\widetilde{\mathbf{X}}_2(\tau)$, we can also determine the limiting vector by choosing specific value of~$ \mathbf{X}_2(0)$ that guarantees~$ \mathbf{X}_2(t)$  to be $T$-periodic. In either way we find the asymptotic vector to be
\bee\label{2-mom-soln-T}
\widetilde{\mathbf{X}}_2(\tau)
 = K_2(\tau,0) Z(T)
 + \int_{0}^{\tau} ds K_2(\tau,s)\mathbf{b}(s)~, 
 \eee
 where
 \bee\label{ZT}
Z(T) = \left[ 1 - K_2(T,0) \right]^{-1} \int_{0}^{T} ds K_2(T,s)\mathbf{b}(s)~.
\eee

To summarize, the asymptotic behavior of the moments of the first two levels, when the condition\eqref{cond-mu-g} holds, is given by the Eqs.\eqref{asym-1} and~\eqref{2-mom-soln-T}.

As an illustrative example, let us consider a simple class of driven harmonic systems, where only~$D(t)$ is taken to be $T$-periodic while~$k$ and~$\gamma$ are kept constant. In this case, the corresponding Hill equation can be explicitly solved and it is straightforward to evaluate~$K_2(t,s)$ from equation\eqref{ker-matrix2} and then find the large time limit of equation\eqref{2-mom-soln} to arrive at the asymptotic moments. The condition\eqref{cond-mu-g} for the existence of the oscillating state reduces to~$k >0$, since~$\gamma >0$. Suppose we expand~$D(t)$, in terms of its Fourier components, as
\bee
D(t) = \sum_{n=-\infty}^{\infty} D_n e^{-i n \omega t}~,
\eee
where~$n$ runs over all integers and~$\omega= 2 \pi/T$. Then we find that these second moments can be written as
\bee
\widetilde{X}_{a,b}(t) = \sum_{n=-\infty}^{\infty} D_n e^{-i n \omega t} \sum_{\sigma=0,\pm 1} \frac{ A_{a,b}^{( \sigma)}} {\gamma + 2 \Omega \sigma -i n \omega}~,
\eee
where~$a$ takes values~$0,1,2$ and~${b=2-a}$, the parameter~$\Omega =\sqrt{ (\gamma/2)^2 - k }$, and
\begin{eqnarray}
A_{a,b}^{( \pm 1)} &=& \frac{1}{4 \Omega^2} \left( \mp \Omega - \frac{\gamma}{2} \right)^b \nonumber ~, \\
A_{a,b}^{(0)} &=&  \frac{-1}{4 \Omega^2} \left[   \left( i \Omega - \frac{\gamma}{2} \right)^b +  \left(- i \Omega - \frac{\gamma}{2} \right)^b \right]~.
\end{eqnarray}
A special case of this simple class, where the only nonzero Fourier components of~$D(t)$ are~$D_0$ and~$D_{1} = D_{-1}$ is a known example\cite{Fiore2019}. The procedure outlined here can be useful in finding explicit solutions in cases where even the parameters~$k$ and~$\gamma$ are driven. 

It should be remarked that time-dependent harmonic oscillators, both classical and quantum, have been extensively studied. In particular, there are numerous studies related to a quadratic dynamical invariant, known as Lewis invariant~\cite{Lewis1968}, with time-dependent coefficients. The dynamics of these coefficients is analogous to Eq.\eqref{x2-eqn} with~$D=0$~\cite{Lewis1969, Korsch1979}.

\subsubsection{The asymptotic distribution}

We now argue that, in the large time limit, the odd level moments vanish while the even level moments satisfy Wick's contraction property.

Note that the level-$L$ moments obey the homogeneous equation\eqref{L-hom} when the level-$(L-2)$ moments vanish. Furthermore,~$\mathbf{X}^h_{L}\tm$ and~$\mathbf{X}_{1}\tm$ vanish for large times, when the condition\eqref{cond-mu-g} holds. Hence, by induction, we conclude that all odd level moments asymptotically vanish.

In case of the even moments, consider the differences~${ \delta_W X_{m,n}= X_{m,n}-X_{m,n}^{(W)} }$ between the moments~$X_{m,n}$ and their corresponding Wick's contracted quantities~$X_{m,n}^{(W)}$. As briefed in the appendix, when these differences up to level-$({L-2})$ moments vanish, the differences~$\delta_W X_{m,n}$ for level-$L$ moments satisfy the homogeneous equation. Further the level-$2$ differences~$\delta_W X_{m,n}$ are identically zero, while level-$4$ ones satisfy the homogeneous equation. We thus conclude, by induction, that the even moments satisfy the Wick's contraction property asymptotically, provided the condition\eqref{cond-mu-g} holds.
 
Hence the asymptotic behavior is essentially governed by the oscillating distribution 
\bee\label{xv-dis}
P_{os}(x,v,t) = \frac{1}{2\pi\sqrt{|\Sigma(t) |}} \mathrm{exp} \left(-\frac{1}{2}
{\begin{bmatrix}
x\\
v
\end{bmatrix} }^T
\Sigma(t)^{-1}
\begin{bmatrix}
x\\
v
\end{bmatrix}
\right)~,
\eee
where~$\Sigma$ is the covariance matrix given by
\bee\label{cov-mat}
\Sigma\tm = 
\begin{bmatrix}
\widetilde{X}_{2,0}\tm & \widetilde{X}_{1,1}\tm\\
\widetilde{X}_{1,1}\tm & \widetilde{X}_{0,2}\tm
\end{bmatrix}~,
\eee
and~$|\Sigma (t)|$ denotes the determinant of~$\Sigma(t)$. The normalizability of the asymptotic distribution, or the existence of the oscillating state, requires~$\Sigma$ to be positive definite.  

The above distribution suggests that the position~$x$ and velocity~$v$ degrees of freedom are in general not decoupled in the oscillating state, unlike the case in the equilibrium state. In other words, the marginal distribution for~$x$ variable obtained from \eqref{xv-dis} is different from the asymptotic distribution of the corresponding overdamped Brownian process. This also suggests that the set of relevant variables required to describe thermodynamic systems can be larger when driven periodically than when not driven.

A few comments, about the rigor in establishing the asymptotic distribution and the requirement to do so, are in order. It is not unusual to consider an ansatz for solving partial differential equations encountered in physical problems. We could consider at the outset, in case of these harmonic systems, a Gaussian ansatz for the asymptotic distribution. Essentially, assume that the expression\eqref{xv-dis} is the solution of the FP equation\eqref{oFP-eqn}, wherein the elements of the covariance matrix\eqref{cov-mat} are to be considered as the parameters of the ansatz. Then the consistency of the assumption requires that these parameters satisfy the ordinary differential equation\eqref{x2-eqn}. Thus the problem of finding the periodic asymptotic distribution reduces to finding the periodic solution of the differential equation\eqref{x2-eqn} involving three variables or, equivalently, a third-order ordinary differential equation. This procedure, though, enables us to bypass elaborate techniques but nevertheless compromises on various counts. A consistent ansatz does not guarantee that it is the asymptotic state. At best we could identify a small domain of attraction for the initial conditions from the study of its stability against small perturbations.
On the other hand, the systematic procedure that we employed here, though requiring sophisticated techniques of representation theory, not only establishes the existence of the oscillating state but also provides us with the condition\eqref{cond-mu-g} required for the solution to exist. This condition can even reveal the existence of an oscillating state in counterintuitive examples of driven systems.
Furthermore, we could show that the distribution is completely deduced once the corresponding Hill equation\eqref{Hill} is solved. In other words, we find an almost exact solution of the FP equation, namely, we find an associated Hill equation, which is just a second-order ordinary differential equation. Hence all the moments of the oscillating state and various thermodynamic quantities can be expressed in terms of the solutions of the Hill equation. We could essentially find the asymptotic distribution almost exactly solely by exploiting the underlying $SL_2$ symmetry or, more precisely, by using the crucial fact that any irreducible representation of the symmetry can be obtained from the symmetrized tensor product of the fundamental $L=1$ representation.

%
%
%
%

\section{Thermodynamic Quantities}\label{3TQ}

In this section, we study various thermodynamic quantities  in the framework of stochastic thermodynamics~\cite{Sekimoto1998, Qian2001, Seifert2005, Sekimoto2010,Seifert2012}.  

Essentially, a stochastic variable~$A_t = A(\mathbf{X}_t)$ is associated to any given observable~$A(\mathbf{x})$, whose distribution is induced by the distribution\eqref{form-sol} of~$\mathbf{X}_t$. In the large time limit, the induced distribution also becomes $T$-periodic, and hence we expect various moments of the thermodynamic observables in the oscillating state to be time dependent and $T$-periodic.

Here we shall only consider the more general underdamped case and evaluate some of the relevant quantities, including energy and entropy averages and their fluctuations in the oscillating state.

\subsection{Energy: Dynamics and fluctuations}
The energy of the Brownian particle in potential~$U(x,t)$ is a stochastic variable, and is expressed as
\bee
E_t := E(X_t, V_t,t) = \frac{1}{2} V_t^2 + U(X_t, t)~.
\eee
The Stratonovich convention is chosen~\cite{Sekimoto1998, Sekimoto2010} for the stochastic dynamics so that the first law of thermodynamics holds strongly, namely,
\bee
dE_t = \dbar Q_t +\dbar W_t ~,
\eee 
where the infinitesimal heat gained by the system
\bee
\dbar Q_t := \left( -\gamma V_t + \eta(t)\right) \circ dX_t =  -\gamma V_t^2 dt + V_t \circ dB_t ~,
\eee
and the infinitesimal work done on the system
\bee
\dbar W_t :=  \frac{\partial}{\partial t } E(X_t,V_t, t) \circ dt = dt \frac{\partial}{\partial t } U(X_t, t) ~.
\eee
The symbol~$\circ$ denotes the Stratonovich product and~${ dB_{t} =\int_t^{t+dt}dt' \eta(t') }$. While determining the expectation of fluctuating quantities, it is convenient to express Stratonovich product~($\circ$) in terms of It\^o product~($\cdot$), for instance,~${ V_t \circ dB_t = V_t \cdot dB_t + D dt }$.

In case of the driven harmonic potential~${ U(x,t)=k(t) x^2/2 }$, the averages of quadratic functions of~$x,v$ in the oscillating state trivially follow from their definition, while the averages of the nonquadratic ones can be evaluated straightforwardly by Wick's contractions. For instance, the rate of work done 
\bee
\frac{\langle \dbar W_t \rangle}{dt} = \frac{1}{2} \frac{ dk(t)}{dt~} \widetilde{X}_{2,0}(t)~,
\eee
the rate of heat dissipation
\bee\label{heat-diss}
\frac{\langle \dbar Q_t \rangle} {dt} = -\gamma(t) \widetilde{X}_{0,2}(t) + D(t)~,
\eee
and the coefficient of heat diffusion
\bee\label{heat-diff}
\frac{\langle \dbar Q_t^2 \rangle}{dt} = 2D(t) \widetilde{X}_{0,2}(t)~.
\eee

We now evaluate the generating function of the moments of the energy in harmonic case. The probability distribution~$P_{os}(E,t)$ of energy~$E_t$ at time~$t$ is given by
\bee\label{E-dis-def}
P_{os}(E,t) = \int_{x,v}~\delta \! \left( E- E(x,v,t) \right) P_{os}(x,v,t)~,
\eee
where~${ E(x,v,t)= (v^2  + k(t) x^2)/ 2 }$ and, for notational simplicity,~$\int_{x,v}$ is used for integration over~$x$ and~$v$. Taking the integral representation of the $\delta$-function and integrating over~$x$ and~$v$ leads to the expression 
\bee\label{E-dis-int}
P_{os}(E,t) = \int_{-\infty}^{+\infty} \frac{d\lambda}{2\pi} \frac{e^{-i\lambda E} }{\sqrt{ \left( 1-i\lambda \langle E_t \rangle \right)^2+\lambda^2  \langle E_t^2 \rangle  }}~,
\eee
Note that the distribution of the energy is completely specified by the first two moments of the energy,~$\langle E_t \rangle$ and~$\langle  E_t^2 \rangle$. It does not explicitly depend on any of the driving parameters.  In the special case, where the parameters are made independent of time, the energy distribution is completely fixed by~$\langle E_t \rangle$, as is expected from equilibrium distribution.

Essentially all the higher moments of energy can be written down in terms of the first two moments. Equivalently, we could consider evaluating the moments of the energy deviation from its mean~${ \Delta E_t := E_t - \langle E_t \rangle }$. The higher moments of the energy deviation can be obtained from the generating function
\bee\label{E-gen-fn}
G_{\epsilon} (t;  \langle E_t \rangle ,  \langle E_t^2 \rangle; \lambda ) =
 \frac{e^{-i\lambda \langle E_t \rangle } }{\sqrt{ \left( 1-i\lambda \langle E_t \rangle \right)^2+\lambda^2  \langle E_t^2 \rangle  } }~,
\eee
by taking derivatives with respect to~$\lambda$. More explicitly, for any non-negative integer~$n$, the corresponding moment can be generated from the expression
\bee\label{E-mom-gen}
\langle \Delta E_t^n \rangle = \left. \left( -i \partial_{\lambda} \right)^n G_{\epsilon} (t;  \langle E_t \rangle ,  \langle E_t^2 \rangle; \lambda ) \right|_{\lambda=0}~,
\eee 
where~$\partial_{\lambda}$ denotes~$\partial /\partial \lambda$.

The averages of energy~$\langle E_t \rangle$ and energy square~$\langle E_t^2 \rangle$ in the oscillating state can be evaluated from the original distribution\eqref{xv-dis}, and can be expressed as
\begin{eqnarray}
\label{E-avg}
&\langle E_t \rangle =  \frac{1}{2} \widetilde{X}_{0,2}(t) +  \frac{1}{2} k(t) \widetilde{X}_{2,0}(t)~, \\
\label{E-sqr-avg}
&\langle E_t^2 \rangle = 3 \langle E_t \rangle^2 - k(t) \left| \Sigma(t) \right|~.
\end{eqnarray}

The energy distribution does not contain complete information about the correlations between~$x$ and~$v$ degrees of freedom. This information can be obtained, for instance, from the combined distribution of kinetic energy~${ E_t^{kin} = V_t^2/2 }$ and the potential energy~${ U_t = U(X_t,t) }$. We simply state without derivation that the combined distribution can be obtained by following similar steps taken for obtaining the energy distribution. Instead we proceed to define a couple of related  quantities that are likely to play a significant role in understanding the properties of the oscillating state. One is related to the average kinetic energy, denoted~$T_s(t)$, which is defined as
\bee\label{eff-temp}
T_s(t) := 2 \langle E_t^{kin} \rangle~,
\eee
and will be referred to as the kinetic temperature of the system.
 The other is the coefficient of correlation~$c(t)$ between the kinetic and the potential energy fluctuations, defined as
 \bee\label{corr-coeff}
 c(t) := \frac {\left\langle \Delta E^{kin}_t \Delta U_t \right\rangle } { \sqrt { \left\langle \left( \Delta E^{kin}_t \right)^2 \right\rangle }\sqrt{ \left\langle \left( \Delta U_t \right)^2 \right\rangle } }~,
 \eee
 This quantity is also a measure of correlation between~$x$ and~$v$.

In case of harmonic potentials, we can easily read the effective temperature~${ T_s(t) = \widetilde{X}_{0,2}(t) }$, and can straightforwardly determine the correlation coefficient
\bee\label{corr-coeff-harm}
c(t) = \frac{\widetilde{X}_{1,1}(t)^2 } {\widetilde{X}_{0,2}(t)\widetilde{X}_{2,0}(t)}~.
\eee 
We shall also find it convenient to rewrite various thermodynamic quantities in terms of~$T_s(t), c(t)$ and~$\langle E_t \rangle$.
For instance, the expression\eqref{E-sqr-avg} can be rewritten as
\bee\label{E-sqr-avg-2}
\langle E_t^2 \rangle = 3 \langle E_t \rangle^2 +(1 \!-\! c(t)) T_s(t)\left( T_s(t) -2 \langle E_t \rangle \right)~.
\eee
In harmonic cases, the equal time distribution has three independent parameters, and hence it is not surprising that we could express the average energy square, or any other observable, in terms of the above three quantities. Equivalently, we could instead choose the quantities~$T_s(t)$,~$c(t)$, and~$\langle U_t \rangle$. Note that there is no equipartition of kinetic and potential energies in the oscillating state, and hence~$T_s(t)$ and~$\langle U_t \rangle$ are in general independent quantities.

\subsection{Entropy: Dynamics and fluctuations}

The entropy of the system in an oscillating state with a given distribution~$P_{os}(x,v,t)$ is also a stochastic variable~$Y_t$, defined  as
\bee\label{S-def}
Y_t := -\ln P_{os}(X_t, V_t,t)~.
\eee
The variable can take any real value~${ y>0 }$. This definition of entropy\cite{Qian2001, Seifert2005} in the stochastic thermodynamics is motivated by the fact that the expectation of~$Y_t$ is the Gibbs entropy:
\bee\label{Y1-mom}
\langle Y_t \rangle = S(t) := - \int_{x,v} P_{os}(x,v,t) \ln P_{os}(x,v,t)~.
\eee

The stochastic process of the entropy~$Y_t$ can be obtained from the expression
\bee
dY_t = - \frac{1}{P_{os}} \left[ \frac{\partial P_{os}}{\partial t}  dt  +  \frac{\partial P_{os}}{\partial X_t} \circ dX_t +  \frac{\partial P_{os}}{\partial V_t} \circ dV_t \right]~,
\eee
where~${ P_{os} \equiv P_{os}(X_t, V_t,t) }$ is used for notational convenience. Using the stochastic dynamics of~$X_t$ and~$V_t$, and the corresponding FP equation, the above expression straightforwardly leads to the relation
\bee\label{Y-process}
dY_t = \frac{\gamma}{D} \dbar Q_t + \left[  \frac{2}{P_{os}} \frac{\partial J_v^{ir}}{\partial V_t}  + \frac{1}{D} \left( \frac{J_v^{ir}}{P_{os}} \right)^2 \right] dt
+ \frac{J_v^{ir}}{DP_{os}} \cdot dB_t ~,
\eee
where~${ J_v^{ir} \equiv J_v^{ir} (X_t, V_t, t) }$ is the irreversible component of the probability current, and is given by the expression
\bee\label{irr-prob-curr}
J_v^{ir} (x,v, t) = - \left( \gamma v  + D \frac {\partial}{\partial v} \right) P_{os}(x,v,t)~.
\eee
Note that the expression\eqref{Y-process} is valid for any potential~$U(x,t)$ and is not just restricted to the harmonic potentials. On the right-hand side of Eq.\eqref{Y-process} the expectation of both the first term in the bracket, provided~${ J_v^{ir} \to 0 }$ as~${ v \to \pm \infty }$, and the last term of the expression vanish. 

Let us assume that the heat bath at any time~$t$  is maintained at an instantaneous temperature~${ T_b(t)=D(t)/\gamma(t) }$, which then allows us to identify the negative of~$\langle \dbar Q_t \rangle /T_b(t)$ as the rate of entropy change of the bath due to the heat transfer. Hence the rate of change of entropy in the system can be expressed as
\bee
\frac{dS(t)}{dt} = \Pi(t) -\Phi(t)~,
\eee
where the rate of entropy production 
\bee\label{ent-prod}
\Pi(t) := \left\langle  \frac{1}{D} \left( \frac{J_v^{ir}}{P_{os}} \right)^2 \right\rangle~,
\eee
and the entropy flux 
\bee\label{ent-flux}
\Phi(t) := -\frac{1}{T_b(t)} \frac{\langle \dbar Q_t \rangle} {dt}~.
\eee
The stochastic quantity~${ dY_t -\dbar Q_t/T_b(t) }$, whose average is~$\Pi(t)dt$ and squared average is~$2\Pi(t)dt$, is essentially associated with the fluctuations of the combined entropy of the system and the bath. 

There has been many recent studies on entropy production~\cite{Tome2010, Spinney2012, Landi2013} including specific periodically driven systems~\cite{Brandner2015, Fiore2019}.
In case of harmonic potential, using Eqs.\eqref{xv-dis},\eqref{cov-mat},\eqref{irr-prob-curr}, and~\eqref{ent-prod}, we can evaluate the rate of entropy production, which can be rewritten as
\begin{eqnarray}\label{ent-prod-har}
\Pi(t)= &\gamma(t) \left[ \frac{T_b(t)}{|\Sigma(t)|}\widetilde{X}_{2,0}(t) + \frac{\widetilde{X}_{0,2}(t)}{T_b(t)} -2\right]  \nonumber \\
= &\gamma(t) \left[ \frac{T_b(t)}{T_s(t) (1-c(t))} + \frac{T_s(t)}{T_b(t)} - 2\right]~.
\end{eqnarray}
Using the identity,~${ r+1/r \ge 2 }$ for any real~${r >0}$, we further conclude that in the oscillating state the entropy production satisfies the inequality,
\bee\label{pi-bound}
\Pi(t) \ge \gamma(t) c(t)  \frac{T_s(t)}{T_b(t)}~,
\eee
and has a nonzero lower bound for any time~$t$, when~${ c(t) >0 }$. When the positivity condition on~$k(t)$ is weakly relaxed at some specific times, without destabiliizng the oscillating state, then the coefficient~$c(t)$ becomes negative and results in no further condition other than the condition~${ \Pi(t) \ge 0 }$, which of course holds at all times.

The entropy flux~$\Phi$, using Eq.\eqref{heat-diss},  can be written as
 \bee\label{ent-flux-har}
\Phi(t) = \gamma(t) \left( \frac{T_s(t) -T_b(t)}{T_b(t)} \right)~.
\eee
Hence, larger the deviation of the system temperature from the bath temperature, then higher is the entropy flux.

 It may be remarked that the bath temperature~$T_b$ is of course related to the three quantities~$T_s(t), c(t)$ and~$\langle U_t \rangle$, though not by a simple algebraic relation. However, in the limit when the potential energy can be neglected, namely as the mass (which is not displayed explicitly here) of the Brownian particle becomes increasingly smaller, then~$T_s$ approaches~$T_b$ increasingly closer.

We now evaluate the moments of the entropy in the oscillation state for harmonic potentials. Substituting Eq.\eqref{xv-dis} into Eq.\eqref{Y1-mom}, and integrating over~$x$ and~$v$, leads to the average entropy
\bee\label{ent-avg}
S(t) = \langle Y_t \rangle = \ln { \left( 2 \pi e \sqrt{\left| \Sigma(t) \right|} \right) }~.
\eee
Similarly we obtain the second moment, which can be written as
\bee\label{ent-2mom}
\langle Y_t^2 \rangle = 1+  \langle Y_t \rangle^2~,
\eee
and find that it is completely fixed by the the first moment alone. Further we note that the dispersion of~$Y_t$ is independent of time.

The probability distribution~$P_{os}(y,t)$ of the entropy~$Y_t$ at time~$t$ is given by
\bee\label{Y-dis-def}
P_{os}(y,t) = \int_{x,v} \delta \! \left( y + \ln{P_{os}(x,v,t)} \right) P_{os}(x,v,t)~.
\eee
Substituting Eq.\eqref{xv-dis} and using the integral representation of the $\delta$-function, in the above expression and then  integrating over~$x$ and~$v$ leads to the distribution 
 \bee\label{Y-dis-int}
P_{os}(y,t) = \int_{-\infty}^{+\infty} \frac{d\lambda}{2\pi} \frac{1 }{ 1-i\lambda } e^{-i\lambda \left( y-   \langle Y_t \rangle +1 \right)}~,
\eee
The probability distribution can also be explicitly written as
\bee
P_{os}(y,t) = \Theta \left( y- \langle Y_t \rangle +1 \right) e^{- \left( y-   \langle Y_t \rangle +1 \right)}
\eee
where~$\Theta$ is the Heaviside step function. The expression\eqref{Y-dis-int} suggests that the moments of the deviation of the entropy from its mean~${ \Delta Y_t := Y_t - \langle Y_t \rangle }$ can be obtained from the generating function
\bee\label{ent-gen-fn}
G_s(\lambda) = \frac{e^{-i \lambda}}{1-i\lambda}~.
\eee
In other words, the moments can be generated from the expression 
\bee\label{Y-mom-gen}
\langle \Delta Y_t^n \rangle = \left. \left( -i \partial_{\lambda} \right)^n G_s ( \lambda ) \right|_{\lambda=0}~.
\eee 
Note that the distribution of the entropy is completely specified by its average~$\langle Y_t \rangle$ and does not explicitly depend on any of the parameters of the system. Though the generating function of~$Y_t$ depends on~$\langle Y_t \rangle$, the generating function~$G_s(\lambda)$ of~$\Delta Y_t$ depend neither on~$\langle Y_t \rangle$ nor on any parameter, and hence is time independent.

\subsection{Energy and entropy correlations}

The energy and the entropy in general can be expected to be correlated even in the oscillating state though the extent of this correlation may not be as much as it is at equilibrium. 

The combined probability distribution~$P_{os}(E,y,t)$ of energy~$E_t$ and entropy~$Y_t$ at time~$t$ is given by
\begin{eqnarray}\label{E-Y-dis-def}
P_{os}(E, y,t) &= \int_{x,v} \delta \! \left( E- E(x,v,t) \right)  \times \nonumber \\
&\times \delta \! \left( y + \ln{P_{os}(x,v,t)} \right) P_{os}(x,v,t)~.
\end{eqnarray}
Following similar steps, as done earlier, we find the energy entropy distribution
\bee\label{E-Y-dis-int}
P_{os}(E, y, t) \!=  \! \int_{\lambda_1, \lambda_2}\! \!\!\! e^{-i \left[ \lambda_1 \left( E \!-\!  \langle E_t \rangle  \right) +  \lambda_2 \left( y \!-\!  \langle Y_t \rangle  \right) \right] } G_{\epsilon,s} (t; \lambda_1,\lambda_2)~,
\eee
where $\int_{\lambda_1, \lambda_2}$ denotes~$\lambda_1,\lambda_2$ integrals over~$(-\infty, +\infty)$ along with the factor~$(2\pi)^{-2}$ and~$ G_{\epsilon,s}(t; \lambda_1,\lambda_2)$ denotes the generating function of the moments of~$\Delta E_t$ and~$\Delta Y_t$, given by
\bee\label{E-Y-gen-fn}
G_{\epsilon,s}(t; \lambda_1,\lambda_2) = \frac{e^{-i\lambda_1 \langle E_t \rangle -i\lambda_2} }{\sqrt{ (1-i\lambda_2 -i\lambda_1 \langle E_t \rangle)^2 +\lambda_1^2  \langle E_t^2 \rangle  } }~.
\eee
Essentially the moments are generated by the expression
\bee\label{E-Y-mom-gen}
\langle \Delta E_t^n \Delta Y_t^m \rangle = \left. ( -i \partial_{\lambda_1} )^n ( -i \partial_{\lambda_2})^m G_{\epsilon,s} (t ; \lambda_1,\lambda_2 ) \right|_{\lambda_1, \lambda_2=0}~,
\eee 
for any non-negative integers~$m$ and~$n$. 

Note that the combined generating function ${ G_{\epsilon,s}(t; \lambda_1,\lambda_2) \equiv  G_{\epsilon,s} (t;  \langle E_t \rangle ,  \langle E_t^2 \rangle; \lambda_1,\lambda_2 ) }$ depends only on~$\langle E_t \rangle$ and~$\langle  E_t^2 \rangle$. The dependence of the moments~$\langle \Delta E_t^n \Delta Y_t^m \rangle$ on~$\langle  E_t^2 \rangle$~, and not just on~$\langle  E_t \rangle$, shows up only when~${ n>1 }$.

 A straightforward analysis leads to the connected correlation function
\bee\label{E-Y-corr}
\langle \Delta E_t \Delta Y_t \rangle = \langle  E_t \rangle~,
\eee
in the oscillating state for the harmonic potentials.
The correlation coefficient~$c_{\epsilon s}(t)$ of the energy and the entropy fluctuations, 
\bee\label{E-Y-coeff}
 c_{\epsilon,s}(t) := \frac {\left\langle \Delta E_t \Delta Y_t \right\rangle } { \sqrt { \left\langle \left( \Delta E_t \right)^2 \right\rangle }\sqrt{ \left\langle \left( \Delta Y_t \right)^2 \right\rangle } }~,
 \eee
can be written, using Eqs.\eqref{E-sqr-avg-2},\eqref{ent-2mom}, and\eqref{E-Y-corr}, as
\bee\label{E-Y-coeff2}
 c_{\epsilon, s}(t) = \left[ 2- 4 (1-c(t)) f_{kin} (1-f_{kin}) \right]^{-\frac{1}{2}}~,
 \eee
where~$f_{kin} = T_s(t)/2  \langle E_t \rangle $ is the fraction of average energy that is kinetic. Note that the correlation coefficient of energy and entropy fluctuations reduces as the correlation between~$x$ and~$v$ increases. The minimum possible value of coefficient~$c_{\epsilon, s}(t)$, when the kinetic and potential energies are positively correlated, is~$1/\sqrt{2}$.
 
In case of equilibrium stochastic thermodynamics the fluctuations of energy~$E_t$ and entropy~$Y_t$ are completely correlated~\cite{Qian2001}, namely, the relation~${ \Delta E_t = T_b(t) \Delta Y_t }$ holds strongly.  While this is not the case in the oscillating state, and the difference~${ \Delta \Psi_t = \Delta E_t - T_b(t) \Delta Y_t }$ is a fluctuating quantity. The generating function~$G_{\psi}(t;\lambda)$ for the moments of~$\Delta \Psi_t$ can be read from the expression\eqref{E-Y-dis-int}, and is given by
\bee\label{psi-gen-fn}
G_{\psi}(t;\lambda) = G_{\epsilon,s}(t; \lambda, -T_b(t) \lambda)~.
\eee
The moments can then be evaluated from the equation
\bee\label{psi-mom}
\langle \Delta \Psi_t^n \rangle = \left. \left( -i \partial_{\lambda} \right)^n G_{\psi} (t;  \lambda ) \right|_{\lambda=0}~.
\eee 
The dispersion of~$\Delta \Psi_t$, using Eqs.\eqref{E-sqr-avg-2},\eqref{ent-2mom}, and\eqref{E-Y-corr}, can be written as
\bee\label{psi-var}
\langle \Delta \Psi_t^2 \rangle =  c(1\!-\!c)T_s^2+ \left( \langle E_t \rangle \! -\! T_b \right)^2  + \left( \langle E_t \rangle \!-\!(1\!-\!c)T_s \right)^2~,
\eee
where the time dependence of~$c(t)$,~$T_s(t)$, and~$T_b(t)$ are not explicitly exhibited for notational simplicity. The first term on the right-hand side of the equation is  due to the presence of correlation between kinetic and potential energies. The second and third terms are due to the deviation of the average energy from~$T_b(t)$ and from the fraction~${ (1-c) }$ of~$T_s(t)$, respectively.

%
%
%
%

\section{Two-time quantities}\label{2Tquant}

In this section, we introduce two-time stochastic variables associated to the correlation and the response functions, and evaluate expectation values of some of these observables. One of the motivations to do so is to study the dynamical response, encoded in the two-time quantities, of the system in the oscillating state. We shall observe that this response can be significantly different from the response of the system at equilibrium.

For any given stochastic variables~${ A_t = A(\mathbf{X}_t) }$ and~${ B_t = B(\mathbf{X}_t) }$, the two-time correlation function in the oscillating state is defined as
\bee\label{2-t-corr}
\left\langle A_t B_{t'} \right\rangle :=\int \! \! d\x d\x' A(\x) B(\x') W_2^{os}(\x,t;\x',t')~,
\eee
where the joint probability distribution
\bee\label{joint-prob}
W_2^{os}(\x,t;\x',t') = P(\x,t|\x',t')P_{os}(\x',t')~.
\eee
can be written in terms of the oscillating distribution~$P_{os}(\x,t)$ and the conditional probability density
\bee\label{cond-prob}
P(\x,t|\x',t') = \mathcal{U}(\x;t,t')\delta(\x-\x')~.
\eee
Using Eq.\eqref{U-Tperiod} in the  above expression leads to identifying the invariance of  the two-time functions under discrete time translation by time period~$T$.

\subsection{Hierarchy of correlation functions}

We first organize all the stochastic variables of the underdamped driven oscillator into a hierarchy of composite stochastic variables. Each hierarchy is labeled by the level index~$L$ and contains only the variables that transform under the~$({L+1})$-dimensional irreducible representation of $sl_2$~algebra. We then show that all the two-time correlation functions can be obtained  when the solution\eqref{sol-Y1} and the equal-time correlation functions are known. 

We denote the level-$L$ composite stochastic variables as
\bee\label{L-r-var}
{\mathcal O}_L^r(t) := X_t^{L-r}V_t^r~,
\eee
where~$L$ is a positive integer and, for any given~$L$, the index~$r$ runs over all the non-negative integers from~$0$ to~$L$. From the dynamics of the stochastic variables~$X_t$ and~$V_t$,  given in Eq.\eqref{stoc-dyn}, we obtain the dynamics of the composite variables as follows
\begin{align}\label{comp-var}
d{\mathcal O}_L^r(t) =& \left[ (L\!-\!r) {\mathcal O}_L^{r+1}(t)- r \gamma {\mathcal O}_L^r(t) - r k {\mathcal O}_L^{r-1}(t) \right] dt
\nonumber  \\  
+& D  r(r-1){\mathcal O}_{L-2}^{r-2}(t) dt + r {\mathcal O}_{L-1}^{r-1}(t) \cdot dB_t~.
\end{align}
The last two terms contain variables only from the lower levels and not from the level-$L$, and the rest of the expression is same as Eq.\eqref{L-hom} with an underlying $sl_2$~symmetry. Hence it follows that these composite stochastic variables can be written as
\begin{eqnarray}\label{L-r-var-2}
{\mathcal O}_L^r(t)&= \left[ K_L(t,t')  \right]^{r~}_{~m}{\mathcal O}_L^{m}(t') 
+  \int_{t'}^t  \left[ K_L(t,s)  \right]^{r~}_{~m} \times \nonumber \\
& \left[ D m(m\!-\!1){\mathcal O}_{L\!-\!2}^{m\!-\!2}(s) ds + m {\mathcal O}_{L\!-\!1}^{m\!-\!1}(s) \cdot dB_s \right]~,
\end{eqnarray}
where the summation over~$m$ from~$0$ to~$L$ is implied, and the $(L\!+\!1)$-dimensional matrix 
\bee\label{funda-matrixL}
K_L(t,s) := \mathbf{\Phi}_{L}(t) \mathbf{\Phi}_{L}^{-1}(s) e^{-\left(\Gamma(t)-\Gamma(s)\right)L/2}~,
 \eee
is formed from the fundamental matrix~$\mathbf{\Phi}_L(t)$ of Eq.\eqref{L-hom2}. The matrix~$\mathbf{\Phi}_L(t)$ can be constructed explicitly, since all the representations of~$sl_2$ can be obtained from the fundamental representation and is essentially the symmetric part of the $L$th~tensor power of~$\mathbf{\Phi}_1(t)$. 

The stochastic variables associated to the two-time correlation functions  are defined as
\bee
C_{LL'}^{rr'}(t,t') ={\mathcal O}_{L}^{r}(t){\mathcal O}_{L'}^{r'}(t')- 
\left\langle {\mathcal O}_{L}^{r}(t) \right\rangle \left\langle {\mathcal O}_{L'}^{r'}(t') \right\rangle~,
\eee
for any pair of levels~${ (L,L') }$ and the pair of components~${ (r,r') }$, respectively. The average two-time correlation function is of the form
\bee
\left\langle C_{LL'}^{rr'}(t,t') \right\rangle = \Theta(t\!-\!t') \overline{C}_{LL'}^{rr'}(t,t')
+ \Theta(t'\!-\!t) \overline{C}_{L'L}^{r'r}(t',t)~,
\eee
since~${ C_{LL'}^{rr'}(t,t')= C_{L'L}^{r'r}(t',t) }$. The functions~$\overline{C}$ then can be evaluated using Eq.\eqref{L-r-var-2},  and we obtain, for~${t>t'}$, the expression 
\begin{eqnarray}\label{corr-fn}
\overline{C}_{LL'}^{rr'}(t,t') = &\left[ K_L(t,t')  \right]^{r~}_{~m} \overline{C}_{LL'}^{mr'}(t')~~~~~~~~~~~~~ \nonumber \\
+  \int_{t'}^t  \left[ K_L(t,s)  \right]^{r~}_{~m} 
 &\left[ D m(m\!-\!1) \overline{C}_{(L\!-\!2)L'}^{(m\!-\!2)r'}(s,t') ds  \right]~,
\end{eqnarray}
where  the equal-time connected correlation functions
\bee\label{conn-corr-1t}
\overline{C}_{LL'}^{rr'}(t) :=  \left\langle{\mathcal O}_{L}^{r}(t)  {\mathcal O}_{L'}^{r'}(t) \right\rangle -
\left\langle{\mathcal O}_L^r(t)  \right\rangle \left\langle{\mathcal O}_{L'}^{r'}(t)  \right\rangle~.
\eee
Thus the two-time correlation functions can be determined iteratively provided equal-time correlations are known.

\subsection{Correlation functions in the oscillating state}
 
We now show that, in the oscillating state, the two-time correlation functions of level-$1$ and level-$2$ variables follow simple relations.

For~$L=1$ and~$L'=1$, Eq.\eqref{corr-fn}, reduces to
\bee\label{K-1-mat}
\overline{C}_{11}(t,t') = K_1(t,t') \overline{C}_{11}(t')~,
\eee
where~$\overline{C}_{11}(t,t') $ are $\overline{C}_{11}(t')$ are the matrices whose components are two-time and equal-time correlation functions, respectively. The above linear relation makes it transparent that the dynamics of~$\overline{C}_{11}(t,t') $ is same as that of $ K_1(t,t') $, with the initial condition~${ \overline{C}_{11}(t',t') =  \overline{C}_{11}(t') }$.

For~${L=2}$ and~$L'=2$, Eq.\eqref{corr-fn} again takes a simple form,
\bee\label{corr-fn-2}
\overline{C}_{22}^{rr'}(t,t') = \left[ K_2(t,t')  \right]^{r~}_{~m} \overline{C}_{22}^{mr'}(t')~.
\eee
Further, since~$K_2(t,t')$ can be written in terms of the elements of~${ K_1(t,t')= \overline{C}_{11}(t,t') \overline{C}_{11}(t')^{-1} }$, the functions~$\overline{C}_{22}^{rr'}(t,t')$  can be expressed in terms of~$\overline{C}_{11}^{rr'}(t,t')$ and level-2 equal-time correlation functions.

From Eqs.\eqref{funda-matrix1},~\eqref{funda-matrix2}, and~\eqref{funda-matrixL}, the matrix~$K_2(t,t')$ can be explicitly obtained and is found to be
\bee \label{K2}
K_2(t,t') = 
\begin{bmatrix}
\kappa_{11}^2 & 2 \kappa_{11} \kappa_{12}& \kappa_{12} ^2\\
\kappa_{11}\kappa_{21}&  \kappa_{11} \kappa_{22} + \kappa_{12}\kappa_{21} & \kappa_{12}\kappa_{22}\\
 \kappa_{21}^2 & 2 \kappa_{21} \kappa_{22} & \kappa_{22}^2
 \end{bmatrix} ~,
\eee
where the elements~${\kappa_{11} = \left[ K_1(t,t') \right]^{0~}_{~0}}$,
${\kappa_{12} = \left[ K_1(t,t') \right]^{0~}_{~1}}$, ${\kappa_{21} = \left[ K_1(t,t') \right]^{1~}_{~0}}$, and~${\kappa_{22} = \left[ K_1(t,t') \right]^{1~}_{~1}}$.
Substituting the above matrix in Eq.\eqref{corr-fn-2}, and using Wick's contractions to rewrite~$\overline{C}_{22}^{rr'}(t')$ in terms of~$\overline{C}_{11}^{rr'}(t')$, will lead us to the following expressions,
\begin{eqnarray}\label{decomp-corr}
\overline{C}_{22}^{00}(t,t')  &=&  2 \left[ \overline{C}_{11}^{00}(t,t')\right]^2  ~, \nonumber \\
\overline{C}_{22}^{01}(t,t') &=&    2  \overline{C}_{11}^{00}(t,t') \overline{C}_{11}^{01}(t,t') ~, \nonumber \\
\overline{C}_{22}^{02}(t,t')  &=& 2 \left[ \overline{C}_{11}^{01}(t,t')\right]^2 ~, \nonumber \\
\overline{C}_{22}^{10}(t,t') &=&     2  \overline{C}_{11}^{00}(t,t') \overline{C}_{11}^{10}(t,t') ~,  \nonumber \\
\overline{C}_{22}^{11}(t,t') &=&     \overline{C}_{11}^{00}(t,t') \overline{C}_{11}^{11}(t,t') +
 \overline{C}_{11}^{01}(t,t') \overline{C}_{11}^{10}(t,t') ~, \nonumber \\
\overline{C}_{22}^{12}(t,t') &=&   2  \overline{C}_{11}^{01}(t,t') \overline{C}_{11}^{11}(t,t')    ~,   \nonumber \\
\overline{C}_{22}^{20}(t,t') &=&     2 \left[ \overline{C}_{11}^{10}(t,t')\right]^2    ~,  \nonumber \\
\overline{C}_{22}^{21}(t,t') &= &     2  \overline{C}_{11}^{10}(t,t') \overline{C}_{11}^{11}(t,t')   ~,   \nonumber \\
\overline{C}_{22}^{22}(t,t') &=& 2 \left[ \overline{C}_{11}^{11}(t,t')\right]^2 ~.
\end{eqnarray}
Note that the above two-time correlation functions do not explicitly depend on the equal-time correlations. Furthermore, they satisfy the Wick's contraction property. 

We can deduce by induction, from the set of equations\eqref{corr-fn}, that the functions~${\overline{C}_{LL'}^{rr'}(t,t')}$ vanish when~${L+L'}$ is not an even integer, which is indeed  a consequence of the symmetry~${ (X_t,V_t) \to (-X_t,-V_t) }$ of the stochastic process. The vanishing of the functions~$\overline{C}_{12}^{rr'}(t,t')$ and~$\overline{C}_{21}^{rr'}(t,t')$ is of course evident, since equal-time correlations~${\overline{C}_{LL'}^{rr'}(t)}$ vanish for odd~$L+L'$.

In general, for any~$L$ and~$L'$, the functions~$\overline{C}_{LL'}^{rr'}(t,t')$ can be determined, though can be a tedious exercise, once we write~$K_L(t,t')$ in terms of the elements of~$K_1(t,t')$.  Essentially, the entire dynamical information in the harmonic case is encoded in the two-time correlation functions~$\overline{C}_{11}^{rr'}(t,t')$. It should be remarked that we could uncover even this dynamical aspect by merely capitalizing on the presence of underlying $SL_2$ symmetry, and thoroughly using the properties of its irreducible representations.  

\subsection{Two-time correlations of energies and entropy}

All the two-time correlation functions involving the stochastic variables~$\Delta E_t^{kin}$,~$\Delta U_t$,~$\Delta E_t$ and~$\Delta Y_t$ can of course be written explicitly, using Eq.\eqref{decomp-corr}, in terms of the four quantities~$\overline{C}_{11}^{rr'}(t,t')$. Equivalently, we can choose any four two-time quantities involving energies, which are more generic observables, and express the rest in terms of them. Let us choose the following correlation functions,
\begin{eqnarray}\label{EE-2time}
&\langle \Delta E_t^{kin} \Delta E_{t'}^{kin} \rangle =  \frac{1}{2} \langle V_t V_{t'} \rangle^2  ~,\nonumber \\
&\langle \Delta E_t^{kin} \Delta U_{t'}\rangle = \frac{1}{2} k(t') \langle V_t X_{t'} \rangle^2 ~, \nonumber \\
&\langle \Delta U_t \Delta E_{t'}^{kin} \rangle = \frac{1}{2}  k(t) \langle X_t V_{t'} \rangle^2  ~,\nonumber \\
&\langle \Delta U_t \Delta U_{t'} \rangle =  \frac{1}{2}k(t) k(t')  \langle X_t X_{t'} \rangle^2 ~.
\end{eqnarray}
which are essentially read off from Eq.\eqref{decomp-corr}.

The entropy deviation~$\Delta Y_t$ in the oscillating state, obtained from Eqs.\eqref{S-def} and\eqref{xv-dis}, is given by
\bee\label{S-deviate}
\Delta Y_t = \frac{1}{2} \Sigma_{22}^{-1}(t) \Delta V_t^2 + \frac{1}{2} \Sigma_{11}^{-1}(t) \Delta X_t^2 + \Sigma_{12}^{-1} \Delta \! \left( X_t V_t \right)~,
\eee
where the notation~${ \Delta A_t := A_t \!-\! \left< A_t \right> }$ is used to denote the deviation of the corresponding variable~$A_t$ from its mean. This expression, on using Eqs.\eqref{cov-mat},\eqref{corr-coeff}, and\eqref{EE-2time},  can be rewritten as
\bee\label{S-deviate2}
\Delta Y_t =\frac{1}{\sqrt{2} \left( 1- c(t) \right) } \left[ \hat{\delta} E_t^{kin} + \hat{\delta} U_t - \sqrt{c(t)} \Delta \Omega_t \right]~,
\eee
where, at least for notational simplicity,  we use a normalized deviation 
\bee\label{nor-var}
\hat{\delta} A_t := \frac{ \Delta A_t} { \sqrt{ \left< \left( \Delta A_t \right)^2 \right>}} ~,
\eee
corresponding to any variable~$A_t$, and the variable
\bee
\Omega_t := \frac{ \sqrt{k(t)} X_t V_t } { \left[ \left< \left( \Delta E_t^{kin} \right)^2 \right>   \left< \left( \Delta U_t \right)^2 \right>   \right]^{\frac{1}{4}}  } ~.
\eee

 The correlation functions involving~$\Delta \Omega_t$ and the energy variables can be straightforwardly determined, by using Wick's contractions and Eq.\eqref{EE-2time}. Thus we obtain the following expressions: 
\begin{eqnarray}\label{R-corr}
\langle \Delta\Omega_t \hat{\delta} E_{t'}^{kin} \rangle &=2 \sqrt{  \langle \hat{\delta} U_t \hat{\delta}  E_{t'}^{kin} \rangle  } \sqrt{ \langle \hat{\delta}  E_t^{kin} \hat{\delta}  E_{t'}^{kin} \rangle }~, \nonumber \\
\langle \Delta \Omega_t \hat{\delta}  U_{t'} \rangle &=2 \sqrt{  \langle \hat{\delta}  U_t \hat{\delta}  U_{t'} \rangle  } \sqrt{ \langle \hat{\delta}  E_t^{kin}\hat{\delta} U_{t'} \rangle }~, \nonumber \\
\langle \Delta \Omega_t \Delta \Omega_{t'} \rangle &= 2 \left[ \sqrt{  \langle \hat{\delta}  U_t\hat{\delta} U_{t'} \rangle  } \sqrt{ \langle \hat{\delta}  E_t^{kin} \hat{\delta}  E_{t'}^{kin} \rangle } \right.\nonumber \\
&+ \left.  \sqrt{  \langle \hat{\delta}  U_t \hat{\delta}  E_{t'}^{kin} \rangle  } \sqrt{ \langle \hat{\delta}  E_t^{kin} \hat{\delta}  U_{t'} \rangle } \right]~.
\end{eqnarray}
The expressions for~$\langle \hat{\delta}  E_{t}^{kin} \Delta \Omega_{t'}  \rangle$ and~$\langle \hat{\delta}  U_{t} \Delta \Omega_{t'} \rangle$ are same as those for~$\langle \Delta \Omega_t \hat{\delta}  E_{t'}^{kin} \rangle$ and~$\langle \Delta \Omega_t \hat{\delta}  U_{t'} \rangle$, respectively, but with~$t$ and~$t'$ interchanged. 

In essence, all the two-time correlations of entropy with energies, or entropy with entropy, are determined by the two-time correlations of kinetic and potential energies. 

In the absence of driving, the correlation~$c(t)$ between kinetic and potential energies vanishes. Hence the presence of $\Omega_t$~term in~$\Delta Y_t$ is purely due to the periodic driving. It is evident, from the expressions~\eqref{R-corr} and\eqref{S-deviate2}, that the effect of periodic driving on the two-time correlations of the entropy with any other observable cannot be obtained, from the corresponding correlation function in the absence of driving, by invoking an effective variable that is independent of the observable. 

\subsection{Response functions}

We now introduce the stochastic variables associated to the linear response of the system to various perturbations. 

Suppose the Markov process\eqref{lan1} is perturbed such that the FP Eq.\eqref{fp1} is modified to 
\bee \label{fp1-pert}
\frac{\partial}{\partial t}P_{\lambda}(\x,t) =  \left[ \mathcal{L}(\x,t)  +  \lambda(t) \delta_{\lambda}\mathcal{L}(\x) \right] P_{\lambda}(\x,t) ~,
\eee
where the time dependent field~$\lambda(t)$ is switched on at time~$t=0$ and the operator~$ \delta_{\lambda}\mathcal{L}(\x)$ is accordingly defined by the perturbation. The solution of the modified FP equation can be formally written as 
\begin{eqnarray}\label{form-sol-mod}
P_{\lambda}(\x,t)= &\mathcal{U}(\x;t,t_0) P_{\lambda}(\x,t_0) ~~~~~~~~~~~~~~~~~~~~\nonumber \\
+& \int_{t_0}^{t} ds \lambda(s) \mathcal{U}(\x;t,s) \delta_{\lambda}\mathcal{L}(\x)P_{\lambda}(\x,s)  ~,
\end{eqnarray}
for any~$t, t_0$. Choosing the initial state~$ P_{\lambda}(\x,0)$, before switching on the field at~$t_0=0$, to be the oscillating state, will reduce the above solution in the linear regime to
\bee\label{form-sol-linear}
P_{\lambda}(\x,t)\!=\! P_{os}(\x,t) \!+\!\! \int_{0}^{t} \!\!ds \lambda(s) \mathcal{U}(\x;t,s) \delta_{\lambda}\mathcal{L}(\x)P_{os}(\x,s)  ~,
\eee
which can be rewritten as
\bee\label{form-sol-linear-2}
P_{\lambda}(\x,t)\!=\! P_{os}(\x,t) \!+\!\! \int_{0}^{t} \!\!ds \lambda(s) \! \!\int \! \!d\x' \Lambda(\x'; s)W_2^{os}(\x,t;\x',s)~,
\eee
where~$W_2^{os}$ is the joint probability distribution, and the function~$\Lambda$ is defined by the expression
\bee\label{res-stoc-var}
\Lambda(\x; t) :=  \left[ P_{os}(\x,t) \right]^{-1} \delta_{\lambda}\mathcal{L}(\x)P_{os}(\x,t) ~.
\eee
Following the standard linear response analysis, the expectation~$\langle A_t \rangle_{\lambda}$ of the stochastic variable~$A_t$ with respect to the perturbed distribution~$P_{\lambda}(\x,t)$ is then given by
\bee
\langle A_t \rangle_{\lambda} = \langle A_t \rangle +  \int_{0}^{t} \!\!ds \lambda(s) 
\left[ \left\langle A_t  \Lambda_s \right\rangle -
\left\langle A_t \right\rangle \left\langle \Lambda_s \right\rangle \right]~,
\eee
where the stochastic variable~$ \Lambda_t = \Lambda(\mathbf{X}_t; t)$ and all the expectations, on the right-hand side, are with respect to the oscillating state. Hence given any stochastic variable~$A_t$ and the perturbation operator~$\delta_{\lambda}\mathcal{L}(\x)$, we can associate to the corresponding response function a two time stochastic variable
\bee
R_{A\lambda}(t,t') :=  \Theta(t \! - \! t') \left[ A_t \Lambda_{t'} - \left\langle A_t \right\rangle \left\langle \Lambda_{t'} \right\rangle \right]~.
\eee

We  can now write down explicitly the stochastic variables associated to the drift and diffusion perturbations. The most general drift perturbations in the underdamped case amounts to modifying the stochastic process specified in Eq.\eqref{stoc-dyn} as follows:
\begin{align}\label{stoc-dyn-pert}
\dot{X}_{t}& = V_{t} ~,\nonumber \\
\dot{V}_{t}& = -\gm \tm  V_{t} - k \tm X_{t} + h^L_r(t) \mathcal{O}_L^{r}(t) + \eta(t)~,
\end{align} 
where~$h^L_r(t)$ are small drift fields switched on at~$t=0$. We associate to each field~$h^L_r(t)$ a  corresponding stochastic variable, 
\bee\label{res-field}
\Lambda_L^r (t) :=  -r \mathcal{O}_{L-1}^{r-1}(t)  + \frac{\gamma}{D} \mathcal{O}_{L+1}^{r+1}(t) +\frac{1}{D}  \mathcal{O}_{L}^{r}(t) \frac{ J_v^{ir}(X_t,V_t,t)} {P_{os}(X_t,V_t,t)}~,
\eee
which is obtained from Eq.\eqref{res-stoc-var}, on using the fact that the above perturbation modifies the FP operator by~$- h^L_r(t)\partial_v \mathcal{O}_L^{r}(t)$, and then substituting Eq.\eqref{irr-prob-curr}. When restricted to driven harmonic potentials, the above expression reduces to 
\bee\label{res-field-har}
\Lambda_L^r (t) :=  -r \mathcal{O}_{L-1}^{r-1}(t)  + \Sigma_{12}^{-1}(t) \mathcal{O}_{L+1}^{r}(t) +\Sigma_{22}^{-1}(t) \mathcal{O}_{L+1}^{r+1}(t) ~.
\eee
Note that~$\left\langle \Lambda_L^r (t) \right\rangle =0$, which is the case for any~$ \delta_{\lambda}\mathcal{L}(\x)$ of the form~$ \partial_{x_{\mu}} \cdots$, and when~$P_{os}(\x,t)$ vanishes at the boundary points of~$x_{\mu}$.

Thus the linear response of~$ \mathcal{O}_L^{r}(t)$ to the perturbation~${h^{L'}_{r'}(t')}$ is captured by the two-time stochastic variable
\begin{eqnarray}\label{res-fn-var}
R^{rr'}_{LL'}(t,t') =  \Theta(t\!-\!t') \left[ -r' C_{L(L'\!-\!1)}^{r(r'\!-\!1)}(t,t')  \right. + ~~~~~~ \nonumber \\
\left.  
 C_{L(L'\!+\!1)}^{r r'}(t,t')  \Sigma_{12}^{-1}(t') +C_{L(L'\!+\!1)}^{r(r'\!+\!1)}(t,t') \Sigma_{22}^{-1}(t') \right].
\end{eqnarray}
Note that the functions~$ \langle R^{rr'}_{LL'}(t,t') \rangle$ vanish for~${L' =L}$, or  more precisely, when~${L+L'}$ is not an odd integer, since~${\overline{C}_{LL'}^{rr'}(t,t')}$ vanish when~${L+L'}$ is not even.

Let us consider the special case where there is a shift in energy~${E_t \to E_t - h(t)X_t  }$, or equivalently, where the field~$h^0_0(t)$ is switched on.  The linear response of the variable~$X_t$ to this drift perturbation, for~${t>t'}$, is given by
\begin{eqnarray}\label{X-res-h}
R^{00}_{10}(t,t') = \Sigma_{22}^{-1}(t') \left< X_t V_{t'} \right>  +  \Sigma_{12}^{-1}(t') \left< X_t X_{t'} \right>  &\nonumber \\
= \frac{ \left< X_t V_{t'} \right> } { T_s(t')}  \frac{1}{1 \!-\! c(t')}\! \left[ 1 \!-\! \sqrt{c(t') \frac{ \langle \hat{\delta}  U_t \hat{\delta}  U_{t'} \rangle  }{ \langle \hat{\delta} U_t \hat{\delta}  E_{t'}^{kin} \rangle  } } \right]&~,
\end{eqnarray}
where the final expression is obtained on using Eq.\eqref{EE-2time}. In absence of periodic driving the above expression of course reduces to~${ \left< X_t V_{t'} \right> / T_s }$. Similarly, the linear response of the variable~$V_t$ to the same drift perturbation, for~${ t > t' }$, is given by
\begin{eqnarray}\label{V-res-h}
R^{10}_{10}(t,t') = \Sigma_{22}^{-1}(t') \left< V_t V_{t'} \right>  +  \Sigma_{12}^{-1}(t') \left< V_t X_{t'} \right> & \nonumber \\
= \frac{ \left< V_t V_{t'} \right> } { T_s(t')}  \frac{1}{1 \!-\! c(t')}\! \left[ 1 \!-\! \sqrt{ c(t') \frac{ \langle \hat{\delta}  E_t^{kin} \hat{\delta}  U_{t'} \rangle  }{ \langle \hat{\delta}  E_t^{kin} \hat{\delta}  E_{t'}^{kin} \rangle  } } \right]&~.
\end{eqnarray}
It is evident from the above two equations that the fluctuation dissipation ratio is not independent of the observable.

The level-2 stochastic variables do not respond to the above linear perturbation, but  instead would respond to the $h_r^1(t)$~fields, as given in Eq.\eqref{res-fn-var}, or to the diffusion perturbations, as discussed below.

The perturbation of the diffusion coefficient~${ D(t) \to  D(t) + \delta D(t) }$  modifies the Fokker-Planck operator by~$\delta D(t)\partial_v^2$, and induces a shift in the expectation~${ \langle  \mathcal{O}_L^{r}(t) \rangle }$ to 
\bee\label{diff-sh}
 \langle  \mathcal{O}_L^{r}(t) \rangle_{\delta D}=  \langle  \mathcal{O}_L^{r}(t) \rangle +  \int_{0}^{t} \!\!ds ~\delta D(s) 
\left\langle   R_L^r(t,s)  \right\rangle~,
\eee
where the response stochastic variable 
\bee\label{resp-diffusion}
R^{r}_{L}(t,t') :=  \Theta(t \! - \! t') \Delta  \mathcal{O}_L^{r}(t) \Delta \Lambda_D(t')~,
\eee
is defined by the responding variable~$\mathcal{O}_L^{r}(t) $, and the probing variable
\bee\label{lamb-diff}
\Lambda_D(t) = 
 \left( \Sigma_{22}^{-1}(t) V_t  +  \Sigma_{12}^{-1}(t) X_t \right)^2 - \Sigma_{22}^{-1}(t)~.
 \eee
The above expression is obtained from Eq.\eqref{res-stoc-var} for~$\delta_{\lambda}\mathcal{L} = \partial_v^2$.
  The response variable can be rewritten in terms of the two-time correlation variables as
\begin{eqnarray}\label{resp-diff-var}
R^{r}_{L}(t,t')& :=  \left[  C_{L2}^{r0}(t,t') \Sigma_{12}^{-2}(t') + C_{L2}^{r2}(t,t')  \Sigma_{22}^{-2}(t')  \right. \nonumber \\
&+ \left.  
 2  C_{L2}^{r 1}(t,t')\Sigma_{12}^{-1}(t') \Sigma_{22}^{-1}(t')  \right]  \Theta(t\!-\!t')~.
 \end{eqnarray}

In order to obtain the linear response of the energies and entropy to diffusion perturbations, it is convenient to recast the stochastic variable~$\Delta \Lambda_D(t)$ in a form that is similar to that of the entropy deviation~$\Delta Y_t$ as given in Eq.\eqref{S-deviate2}. By following the similar steps taken there, Eq.\eqref{lamb-diff} leads to the expression
\bee\label{lam-ee}
\Delta \Lambda_D (t) \!=\!\frac{\sqrt{2}}{ T_s(t) \! \left( 1\!-\!c(t) \right)^2 }\!\! \left[ \hat{\delta}  E_t^{kin} \!+\!c(t) \hat{\delta}  U_t \!-\! \sqrt{c(t)} \Delta \Omega_t \right]~,
\eee
which can be rewritten, using Eq.\eqref{S-deviate2}, as
\bee\label{lam-ee2}
\Delta \Lambda_D (t) \!=\!\frac{\sqrt{2}}{ T_s(t) \! \left( 1\!-\!c(t) \right) }\!\! \left[ \sqrt{2}\Delta Y_t - \hat{\delta}  U_t  \right]~.
\eee
Hence the linear response of any stochastic variable to diffusion perturbations  in the presence of periodic driving is same as that in the absence of the driving up to an amplitude modulation that is independent of the responding observable. In other words, the explicit relations of the response functions, when written in terms of correlation functions of the responding variable with entropy and with potential energy, remain the same even in the driven case provided the kinetic temperature~$T_s$ is replaced by an effective temperature~${ T_s(t) \left( 1\!-\!c(t) \right) }$ that is $T$-periodic.

\section{Conclusion}\label{conc}

To summarize, we have obtained the asymptotic states and studied the thermodynamic properties of driven harmonic Langevin systems. Under certain conditions these asymptotic states, referred to as oscillating states, exist and can be described by time-periodic distributions.

We notice that the oscillating states can sustain even when the negative semidefinite property of the FP operator is relaxed at times.
We have demonstrated this point explicitly in case of driven overdamped Brownian particle in a time-dependent harmonic potential.

We largely studied the asymptotic properties of driven underdamped Brownian particle in harmonic potentials.
 We exploited the underlying $SL_2$~symmetry and certain features of its irreducible representations to obtain the asymptotic distribution. 
 We organized the moments and various other stochastic observables based on their transformation properties under the $SL_2$~symmetry.
 
 We have analyzed various thermodynamic quantities including energies and entropy in the oscillating state. These quantities and their fluctuations could be expressed in terms of the kinetic temperature, average energy (or average potential energy), and the correlation coefficient of kinetic and potential energies. We also find that the energy entropy correlations in the oscillating state are quite different from those at equilibrium. We further notice that there is a lower bound on the entropy production in the oscillating state proportional to the kinetic temperature and the correlation coefficient.
 
 We have also analyzed two-time correlation functions in the oscillating state.  The entire dynamical information in the harmonic case is essentially encoded in the level-1  two-time correlation functions, or equivalently, in the two-time correlation functions of kinetic and potential energies. We have shown that all the higher level two-time correlation functions can be determined iteratively. 
  
 We briefly studied the response of the stochastic system to drift and diffusion perturbations. The response of any variable to the diffusion perturbation could be written in terms of correlation functions of that variable with entropy and with potential energy. These relations are found to remain the same even in the presence of driving, provided the kinetic temperature is replaced by a $T$-periodic effective temperature.  Fluctuations of the two-time stochastic variables can also be studied by the methods employed here. 
 
 Some of the relations obtained  here are of course specific to harmonic potentials.  It would be interesting to ask whether any of these relations survive when more general potentials are considered.  
 
Our analysis can be easily extended to many particle driven stochastic systems with harmonic interactions. Non-harmonic interactions can also be included, if treated perturbatively. The systematic nature of the analysis further provides the motivation to try applying similar methods to stochastic systems with other symmetries.

\appendix

\section{Wick's contraction property}

The Wick's contracted quantity~$X_{m,n}^{(W)}$, corresponding to the moment~$X_{m,n}$ of any level~${L=m+n}$, is defined as the sum of products of second moments and is given by
\begin{equation}\label{XW-def}
X_{m,n}^{(W)} = \sum_{\text{all pairs}} \langle \bar{x}_{i_{1}} \bar{x}_{i_{2}} \rangle \langle \bar{x}_{i_{3}} \bar{x}_{i_{4}} \rangle \dots\langle \bar{x}_{i_{L-1}} \bar{x}_{i_{L}} \rangle 
\end{equation}
where the summation is over all possible pairings of second moments, and the~$L$ number of variables~$\bar{x}_i$ denote either~$x$ or~$v$  which appear exactly~$m$ and~$n$ times, respectively. It is implied from the above definition that the level~${L=m+n}$ is even.
When the difference between the moment and its Wick's contracted quantity
\bee\label{delW}
\delta_W X_{m,n}= X_{m,n} -X_{m,n}^{(W)} ~,
\eee
vanishes, then we say that the Wick's contraction property holds for the moment~$X_{m,n}$. 

We can of course write down the dynamical equation for the difference~$\delta_W X_{m,n}$ using Eq.\eqref{om2}. To this end, instead of directly using the expression\eqref{XW-def}, we find it convenient to first decompose~$X_{m,n}^{(W)}$ of level~$L$ in terms of level~${(L\!-\!2)}$ contracted quantities and second moments. This decomposition though is not unique and, for instance, can be chosen as 
\begin{equation}\label{xcon}
X_{m,n}^{(W)} = (m-1) X_{2,0} X_{m-2,n}^{(W)} + n X_{1,1} X_{m-1,n-1}^{(W)}~,
\end{equation}
when~$m \ge 1$. Another convenient choice, when~$n \ge 1$, is
\begin{equation}\label{vcon}
X_{m,n}^{(W)} = (n-1) X_{0,2} X_{m,n-2}^{(W)} + m X_{1,1} X_{m-1,n-1}^{(W)}~.
\end{equation} 
Nonuniqueness of the decomposition can also lead to various algebraic relations. For example, when~$m \ge 1$ and~$n \ge 1$, Eqs.\eqref{xcon} and\eqref{vcon} lead to the relation
\begin{eqnarray}\label{xvrel}
& (m - 1) X_{2,0} X_{m\!-\!2,n}^{(W)} + (n-m) X_{1,1} X_{m\!-\!1,n\!-\!1}^{(W)} \nonumber \\
& - (n-1) X_{0,2} X_{m,n\!-\!2}^{(W)} = 0~.
\end{eqnarray}
Based on the above expression, for later notational convenience, we define the quantity 
\begin{eqnarray}\label{xvrel2}
&\mathcal{C}_{m,n} := (m - 1) X_{2,0} X_{m\!-\!2,n} + (n-m) X_{1,1} X_{m\!-\!1,n\!-\!1} \nonumber \\
& - (n-1) X_{0,2} X_{m,n\!-\!2}~,
\end{eqnarray}
for~$m \ge 1$ and~$n \ge 1$.

We now show that if the Wick's contraction property holds to level-$(L\!-\!2)$ moments, then the differences~${ \delta_{W} X_{m,n} }$ for the level-$L$ moments will satisfy the homogeneous equation\eqref{L-hom}. 
We shall use either Eq.\eqref{xcon} or Eq.\eqref{vcon} for~$X_{m,n}^{(W)}$ at level~$L$ and at level~${L-2}$, depending on the values of~$m$ and~$n$, and assume that the Wick's contraction property holds, namely~$X_{m,n} = X_{m,n}^{(W)}$, at level~${L-2}$ and at level~${L-4}$. Then the time derivative of~$\delta_W X_{m,n}$ can be obtained from Eq.\eqref{om2}, by elementary algebraic manipulation though cumbersome, and we find that the resulting expression takes the simple form
\begin{eqnarray}
\frac{d}{dt}\delta_{W} \!X_{m,n} =&m \delta_{W} \!X_{m\!-\!1,n\!+\!1}
\!-\!n \gamma \delta_{W}\! X_{m,n} \!-\! n k \delta_{W}\! X_{m\!-\!1,n\!+\!1}  \nonumber \\
& + \mathcal{S}_{m,n}  ~,
\end{eqnarray}
where
\begin{eqnarray}
\mathcal{S}_{m,n} =  \begin{cases}
\mathcal{C}_{m-1, n+1}~, \text{ for  } m \ge 2 \text{ and } n \ge 0~, \\ 
k \mathcal{C}_{m+1, n-1}~, \text{ for  } n \ge 2 \text{ and } m \ge 0 ~,
\end{cases}
\end{eqnarray}
obtained by the choice\eqref{xcon} and the choice\eqref{vcon}, respectively.
Hence we conclude, using Eq.\eqref{xvrel}, that the dynamics of level-$L$ differences~$\delta_{W} \!X_{m,n}$ is given by the homogeneous Eq.\eqref{L-hom}. Note that the Wick's contraction property trivially holds for~$L=2$ case.

For level~$L=4$, using the original defining Eq.\eqref{XW-def}, it is easy to verify that the differences~$\delta_{W} \!X_{m,n}$ satisfy the homogeneous equation\eqref{L-hom}. Hence the Wick's contraction property holds asymptotically for~$\delta_{W} \!X_{m,n}$ at level~$4$ and by induction at all even levels, provided the condition\eqref{cond-mu-g} is satisfied.

\end{document}